# Tailoring Heterovalent Interface Formation with Light

Kwangwook Park[1] and Kirstin Alberi[1,] *

[1]National Renewable Energy Laboratory, Golden, Colorado, 80401, United States
*kirstin.alberi@nrel.gov

**Abstract**

Integrating different semiconductor materials into an epitaxial device structure offers additional degrees of freedom to select for optimal material properties in each layer. However, interface between materials with different valences (i.e. III-V, II-VI and IV semiconductors) can be difficult to form with high quality. Using ZnSe/GaAs as a model system, we explore the use of UV illumination during heterovalent interface growth by molecular beam epitaxy as a way to modify the interface properties. We find that UV illumination alters the mixture of chemical bonds at the interface, permitting the formation of Ga-Se bonds that help to passivate the underlying GaAs layer. Illumination also helps to reduce defects in the ZnSe epilayer. These results suggest that moderate UV illumination during growth may be used as a way to improve the optical properties of both the GaAs and ZnSe layers on either side of the interface.

## Introduction

Various semiconductor materials are commonly incorporated into semiconductor devices to meet their performance needs. Material selection is made on the basis of electronic structure and optical and electrical properties, although consideration is typically given to matching crystal structure or thermal expansion coefficients to allow for monolithic growth in a single epitaxial stack. More often than not, these materials are restricted to specific classes of semiconductors, such as group IV, III-V or II-VI semiconductors, for ease of growth. However, there are obvious advantages for incorporating materials from several classes to select for very specific properties in each layer and/or the substrate. This includes the growth of III-V optoelectronic devices on Si substrates and the application of high bandgap II-VI cladding layers on III-V absorbers or light-emitting active layers. Combining dissimilar materials therefore creates myriad possibilities for designing next generation semiconductor devices if the integration can be carried out successfully.

Interfaces are an inevitable product of material integration and the formation of heterovalent interfaces presents a particular difficulty for growth. Achieving the correct distribution of bonds across the interface while minimizing defect formation requires precise control of the interface initiation conditions. For example, point defect and secondary phase formation and atomic diffusion between layers can occur during the growth of a ZnSe/GaAs hetero-structure by molecular beam epitaxy (MBE) if the interface is not initiated properly [20]. Such defects would be detrimental to device performance if the heterovalent interface is located between critical layers.

Two free parameters that are typically available during growth to provide this control are temperature and molecular or atomic flux. Temperature is the easiest to adjust and strongly influences most growth processes. High growth temperatures are advantageous for enhancing adatom surface mobility and producing high quality crystalline epilayers. Yet, elevated temperatures can also promote atomic inter-diffusion between adjacent layers, which can degrade their optical and transport properties. It may therefore be beneficial to use lower growth temperatures to achieve abrupt interfaces [1]. Low temperatures can additionally lead to more abrupt doping profiles [2] and extend the critical thickness of lattice mismatched layers [3], but they can also inhibit adatom surface migration, promote three-dimensional growth and produce higher defect concentrations [4, 5]. The elemental source flux ratio may also be adjusted to modify surface migration, defect formation and the bonding environment at the interface, although care must be taken to avoid unwanted micro- and nano-structure formation within the epilayer [6, 29]. Optimal control of growth processes therefore may not be achieved by varying these two parameters alone.

Light stimulation of the growth surface offers an additional parameter with which to alter growth processes [7-15, 28]. In particular, it has been shown to improve the crystal quality of II-VI semiconductor epilayers grown at temperatures as low as 150°C and has helped to improve extrinsic doping efficiency [16, 17]. This approach appears to be advantageous for forming heterovalent interfaces because it permits the use of low growth temperatures without sacrificing material quality. Nevertheless, most research has focused on the effect of light stimulation on the II-VI epilayer only, and very little attention has been paid to the effect of light on the initiation of a II-VI/III-V interface or its affect on the underlying III-V material.

Herein, we present an investigation into the effect of light stimulation during low temperature growth of II-VI/III-V interfaces, using ZnSe/GaAs as a model system. Its effects on the optical, structural and morphological properties of both epilayers as well as the interface were studied as a function of illumination conditions. We find that UV illumination alters the elemental coverage of the surface at the start of the initiation process, leading to a different mixture of bonds at the interface. This results in an apparent improvement in the passivation of the underlying GaAs epilayer. UV illumination is also improved the optical properties of the ZnSe epilayer. Based on these findings, we offer some guidance for using light-stimulated growth techniques for heterovalent interface formation.

## Result and discussion

ZnSe epilayers were grown on GaAs epilayers, which were both grown by MBE on GaAs substrates. Initiation of the ZnSe/GaAs interface commenced with a Zn pre-exposure of an As-terminated GaAs surface, followed by growth of the ZnSe layer. Details of the growth are included in the Methods section below and are illustrated in Fig. 1. A Xe lamp was used as the illumination source, and the illumination conditions during the interface initiation process were varied to determine the mechanisms by which UV illumination modifies its properties. We investigated three scenarios: 1) no light exposure (referred to as "dark reference"), 2) light exposure starting after the Zn pre-treatment (referred to as "dark-start" and 3) light exposure starting before the Zn exposure phase (referred to as "light-start").

*Structural Properties*

Figure 2(a)-(b) shows ω-2θ HRXRD curves of light-start and dark-start samples, respectively. The thicknesses of ZnSe layers were the same regardless of the starting condition or the lamp power. The primary difference between the two starting conditions is the appearance of weaker Pendellösung fringes in the HRXRD curves of the light-start samples for a given lamp power. Among the light-start samples, the fringes also strongly weaken with increasing Xe lamp power. Disappearance of the fringes could be linked to two possibilities; degradation of the ZnSe epilayer crystallinity or compositional grading at the ZnSe/GaAs interface. SEM images were taken to investigate the prospect of the latter effect. Figure 2(c) and 2(d) shows tilted SEM images of the light-start sample exposed to a lamp power of 150 W and a dark reference sample.

The hemispheric objects on the sample surfaces are SeO$_2$ islands [18]. Both samples exhibited a change from bright to dark contrast from top to bottom indicating a transition from the ZnSe epilayer to the GaAs buffer and substrate. First derivatives of the image contrast profile through the sample thicknesses were obtained to more clearly identify the ZnSe/GaAs interface. The contrast profile was abrupt in dark reference sample but was graded at the interface of the light-start sample. Additionally, the surfaces of both samples were quite smooth. Together, these results indicate that the degradation of the Pendellösung fringes in the HRXRD curves is due to compositional grading at the ZnSe/GaAs interface. Higher lamp power and light-start conditions both appear to enhance the grading effect. It is noteworthy that the growth temperature used here (200°C) is far below the optimal ZnSe growth temperature (300°C) and in general, lower growth temperature is advantageous for creating abrupt ZnSe/GaAs interfaces. Thus, the role of illumination before and during the Zn pre-treatment phase is somewhat similar to increasing the substrate temperature, which promotes more interface intermixing. Excessive atomic intermixing at the interface was also observed by our group when growing ZnSe epilayers on As-depleted GaAs surfaces [20]. In reference [20], the width of the ZnSe HRXRD peak broadens and the Pendellösung fringes between the ZnSe and GaAs peaks disappears with decreasing As coverage. Depletion of the As coverage at the start of the interface initiation allows more intermixing of Ga, Zn and Se. UV illumination is known to selectively enhance anion desorption from the surface via disruption of the anion bonds by photo-generated holes and has been observed specifically to promote As desorption from GaAs surfaces by our group [22]. The HRXRD results therefore suggest that UV photon irradiation acts to deplete the As coverage of the GaAs starting surface in a similar manner as elevated substrate temperatures. On the other hand, applying the pre-treatment steps before exposing the surface to UV photons appears to set the bonds at the interface in place and prevents substantial intermixing. The Pendellösung fringes remained strong in the HRXRD curves of all dark-start samples, regardless of the lamp power. This result indicates that a Zn sub-monolayer stabilizes the interface surface to some degree, and subsequent exposure to light does not substantially affect it.

*Morphological Properties*

Figure 3 shows AFM images of the light-start samples and their corresponding power spectral densities (PSD). The PSDs provide information about the frequencies of spatial feature wavelengths. The general form of the PSD is derived from continuum growth models and is expressed as $\delta_t h \propto \nabla^2 h + \nabla^4 h + \cdots$ [21]. To produce PSD curves in a wide range of spatial frequencies, AFM images were obtained over several areas: 1×1 μm$^2$, 5×5 μm$^2$ (not shown here), and 10×10 μm$^2$. Care was taken not to include features from the SeO$_2$ phases in the analysis. As shown in Figure 3(a)-(d), the PSD curve slope of the dark reference sample falls between 2 nm$^3$/um$^{-1}$ and 4 nm$^3$/um$^{-1}$. This slope moves closer to 2 nm$^3$/um$^{-1}$ with increasing lamp power as shown in Figure 4(e), indicating that the surface features become somewhat larger. There is an additional feature in the PSD spectra, marked with boxes in Figures 3(a)-(d), that represents the periodicity of hillocks in the range of 0.1 - 1 μm (1 - 5 μm$^{-1}$). In the dark reference sample, the two regions of elevated intensity represent the prevalence of two surface hillocks having periodicities of 0.5 μm (2 μm$^{-1}$) and 0.2 μm (5 μm$^{-1}$). An increase in the light intensity broadens and lowers the range of dominant spatial frequencies, indicating that the surface hillocks have generally become wider (i.e. there is a shift to lower spatial frequencies). We should note that the actual surface features associated with the ZnSe epilayer are never more than a few nanometers in height. The substrate temperature and elemental coverage (i.e. As, Se, etc) strongly affect the growth mode. Low temperatures tend to reduce adatom mobility, leading to 3D growth modes and roughening. This is the case in the present situation, where the ZnSe epilayers were grown well below the optimal growth temperature (200°C instead of 300°C). Standard growth of ZnSe also relies on a high Se/Zn flux ratio, which leaves the surface Se terminated. While UV illumination enhances anion desorption, the cation desorption rate is less affected [23, 26]. The resulting surface becomes more metal rich, leading to improved adatom mobility and migration and to smoother surfaces [19]. This appears to be a plausible explanation for the broadening of the surface features. Figure 4 shows surface AFM images of the dark-start samples and their corresponding PSD curves. The behavior is much the same as that of the light-start samples. For a given lamp power, there is a slightly greater decrease in the PSD slope compared to the light-start samples, shown in Figure 4(e), indicating a smoother surface with wider hillock structures. This could be due to a decrease in any roughening that can occur when the interface itself is grown under UV illumination. However, the effect appears to be minimal.

*Optical Properties*

Figures 5(a)-(f) shows the PL spectra of the ZnSe epilayers in the light-start and dark-start samples. The emission spectra can be divided into two important regions: near-band-edge emission (NBE) around 2.8 eV and deep-level emission (DLE) below 2.6 eV. Within the DLE region, emission around 1.95 eV is usually referred to as the self-activated (SA) region and has been linked to distant donor acceptor pair transitions, especially those involving zinc vacancies ($V_{Zn}$) and group III donor impurities [24, 30, 27]. The NBE region consists of free and donor-bound exciton emission (FX and DX) peaks as well as a deep acceptor-bound exciton ($I_1^{deep}$) emission and its phonon replicas. The overall trends in the light-start and dark-start samples are similar. The total emission intensity increases when grown under UV irradiation compared to the dark-reference sample. This includes emission in the NBE and DLE regions, and the intensity rise is generally attributed to a reduction in non-radiative recombination. The ratios of the intensities of the peaks in the NBE and DLE regions are plotted in Figure 5(g) for both sets of samples.

The evolution of the SA emission as a function of lamp power is most evident. It first increases upon irradiation with a lamp power of 80 W and then subsequently decreases with increasing lamp power up to 150 W. It is unclear whether the slight increase at 80 W is due to an increase in the $V_{Zn}$ or group III (possibly Ga) impurity concentration or a change in the transfer rate of carriers to those states, which could enhance emission. Overall, the NBE/DLE emission does not change significantly as the lamp power is increased from 0 W to 80 W, which suggests that most of the SA emission increase is due to an overall increase in the number of carriers that reach those acceptor and donor states, rather than the density of states themselves. What is clear is that the NBE/DLE ratio substantially increases as the lamp power is further increased to 150 W. That increase is largely caused by a reduction in the SA emission rather than an increase in the NBE. As noted above, Se desorption is known to increase under UV irradiation, leading to the possibility of forming a more metal-rich growth surface. A greater amount of Zn on the growth surface would indeed reduce the concentration of $V_{Zn}$ in addition to promoting smoother surfaces. To provide a quick check of this analysis, we compared the evolution of the SA emission to another instance where Se desorption is enhanced: elevated substrate temperatures. The PL spectra of two samples grown under dark conditions at 200°C and 300°C are shown in Figure 5(h). The sample grown at 300°C also exhibits a reduction in the SA emission compared to the dark-reference sample grown at 200°C. The progression of the elemental surface coverage is also responsible for a similar trend in the $I_1^{deep}$ intensity, which is also associated with Zn vacancies [25].

Strong UV photon irradiation also leads to an increase in the DX and FX emission in the NBE region. In particular, the emission shifts from a dominant DX peak to a dominant FX peak. This is correlated with a decrease in DLE and an overall improvement in material quality. The NBE/DLE ratio, which can serve here as a proxy for ZnSe epilayer quality, becomes comparable to that of ZnSe grown in the dark at 300°C. This is approximately a factor of 4 times the ratio of the dark reference sample.

We note here that the effect of UV irradiation on samples grown with a light-start or a dark-start are similar, meaning that the improvement is associated with the bulk ZnSe epilayer. The slightly higher NBE/DLE ratio in the dark-start sample grown under a lamp power of 150 W relative to the light-start sample may be due differences in bonding at the interface. However, there is not enough evidence to suggest this is a statistically significant difference between light-start and dark-start conditions.

Further information about the effect of photon irradiation on the ZnSe/GaAs interface can be obtained from PL measurements of the underlying GaAs, shown in Figure 6. The PL spectra of light-start and dark-start samples as a function of illumination conditions are shown in Figures 6(a)-(f), and they exhibit three features: DX and FX emission near 1.51 eV, carbon-bound ($C_{As}$) emission at 1.49 eV, and a broad emission peak at 1.48 eV that exhibits a low energy tail. The emission peak at 1.48 eV is related to Ga-Se bonding at the interface, and it evolves with illumination conditions [32]. In particular, the intensity of this peak increases substantially when interface initiation is carried out with a light-start at the highest lamp power. The HRXRD measurements show evidence of more elemental intermixing at the interface caused by excessive As desorption under these start conditions, which could account for additional Ga-Se bond formation. To confirm the origin of this behavior, we simulated conditions of excess As desorption by increasing the As de-capping temperature prior to interface initiation. Figure 5(g) shows the evolution of the PL peak at 1.48 eV when the As de-cap process was carried out at elevated temperatures up to 670 °C, and the ZnSe epilayer was subsequently grown at 200 °C under dark conditions. This peak becomes evident at desorption temperatures of 500 °C and overwhelms the $C_{As}$ peak when As de-capping is carried out at 670 °C. The intensity of the 1.48 eV peak in the light-start/150 W sample falls somewhere in between these two samples, strongly supporting our claim that the 1.48 eV emission is enhanced by additional As desorption.

Besides changing the amount of As coverage on the starting GaAs surface prior to interface formation, elevated ZnSe growth temperatures can also promote interface intermixing, leading to similar results. The PL spectra of two ZnSe epilayers grown under dark conditions at 200 °C and 300 °C are shown in Figure 6(h). The sample grown at 300 °C also exhibits a moderate PL peak at 1.48 eV, suggesting that Ga-Se bonds have formed across the interface at higher concentrations than in the sample grown at 200 °C.

Comparatively, the dark-start samples do not show such an enhancement in the 1.48 eV peak because the dark-start protects against As-desorption. In Figure 6(d), this portion of the spectrum has been multiplied to show that there is a weak peak near 1.48 eV, but its overall intensity is much lower than the light-start sample grown under similar conditions. This behavior arises through the formation of a abrupt interface upon Zn pre-treatment of an As-rich surface, preventing excessive Ga-Se bonding.

The DX and FX emission of the light-start samples is also enhanced with increasing UV lamp power. This effect is likely the result of lower interface recombination, possibly by Se passivation [31] or a stable interface formed through Ga-Se bonds [20, 32]. Based on this result, a light-start appears to help enhance the underlying GaAs epilayer emission. Excitonic emission of dark-start samples, however, is comparable to or lower than that of the dark reference sample. It is possible that a greater percentage of As-Zn bonds increases interface recombination.

It is also noteworthy that $C_{As}$ acceptor bound excitonic emission is largely suppressed in the dark-start samples, which it is present in both the dark reference at light-start samples. This effect is highly reproducible, and we have observed it in many additional sample sets not detailed here. SIMS measurements, shown in Figure 7, indicate that there is no significant difference in the C concentration at the ZnSe/GaAs interface, or within the ZnSe epilayer and GaAs buffer. We do not understand the origin of this effect, but we currently suspect that it is due to some mechanism that alters exciton binding to C acceptors.

*Observations*

When considered together, the structural, morphological and optical results presented here suggest that UV photon irradiation of the ZnSe/GaAs interface during its formation (light-start conditions) acts to deplete the As surface coverage, leading to a different bonding profile and a very slight roughening of the surface compared to dark initiation conditions. While there is evidence of more atomic intermixing at the interface in the light-start samples from HRXRD measurements, the PL results indicate that a light-start under moderate lamp power (115 W or 150 W here) may be somewhat beneficial for reducing recombination at the interface and improving excitonic emission within the GaAs layer. Similar results may be achieved simply by depleting the As coverage of the GaAs buffer prior to interface initiation through thermal rather than optical means, as shown in Figure 6(g). However, this would require very precise control of the temperature and time of the desorption process to achieve a favorable As coverage, as demonstrated in our previous work [20]. Thus, light appears to provide the right amount of regulation, allowing control over other growth parameters to be relaxed. A dark-start interface initiation approach, on the other hand, produces a much more abrupt ZnSe/GaAs interface and suppresses $C_{As}$ recombination in the GaAs. Thus, a dark-start may be advantageous from this perspective.

In either case, UV photon irradiation during ZnSe epilayer growth at 200°C is instrumental in improving its quality. In fact, the PL spectra of the ZnSe epilayers grown under UV illumination at 200°C were similar to those grown under dark conditions at 300°C, as shown in Figure 5(h), with both exhibiting high FX and DX emission and low DLE and $I_1^{deep}$ emission. Growth of the ZnSe epilayer at 200 °C further prevents interface intermixing. Overall, it appears that growth at 200°C using a Zn pre-treatment and a light-start provides a balance of improved luminescence from both the GaAs and ZnSe. This is important in applications where the light emission properties of the underlying III-V semiconductor are important to device operation (i.e. in applications where the II-VI layer is used to clad the III-V layer). More broadly, irradiation of the growth surface with above-bandgap photons may be a useful tool for modifying the structure or chemical composition of a variety of interfaces.

## Conclusion

In summary, ZnSe epilayers were grown at 200°C on As-terminated and Zn pre-treated GaAs surfaces under an array of UV photon irradiation conditions. HRXRD, SEM, AFM, SIMS, and PL measurements reveal that photon irradiation during interface initiation (light-start) enhances As desorption and allows additional Ga-Se bonding to occur at the ZnSe/GaAs interface. This approach led to some improvement in the emission from the GaAs buffer. Under both light-start and dark-start conditions, light-stimulation substantially improved the ZnSe epilayer material. These results suggest that low temperature growth accompanied by moderate UV photon irradiation can improve the properties of the ZnSe/GaAs interface such that both materials exhibit high optical quality.

## Methods

*Sample Growth*

Samples were grown on semi-insulating (100) GaAs substrates in an Omicron EVO25 MBE system with separate III-V and II-VI semiconductor growth chambers. The two chambers were equipped with conventional K-cells for group-III and/or group-II elemental sources and valved cracker effusion cells for the group-V and group-VI elemental sources. The substrate temperature was measured with a k-Space BandiT band-edge thermometry system with an accuracy of ±1.0°C. The substrates were first outgassed at 300°C in an ultrahigh vacuum chamber before being transferred to the III-V growth chamber, where the oxide layer was thermally desorbed at 610°C for 10 minutes under an $As_2$ overpressure. A 500 nm-thick homoepitaxial GaAs buffer layer was then grown, followed by growth of a thick amorphous As layer at room temperature for one hour which was used to protect the GaAs surface while the sample was transferred to the II-VI growth chamber. Once inside the II-VI chamber, the amorphous As cap was desorbed at 330°C in the absence of any group II or VI overpressure to leave a partially As-terminated surface [20]. In a few specified cases, the As cap desorption temperature was varied. Immediately after As desorption, the substrate temperature was set to 200°C. Interface initiation commenced with exposure of the As-terminated surface to a Zn flux for 2 minutes to form a sub-monolayer (ML) of Zn. This was followed by a 30 seconds exposure to Se, after which the Zn flux was turned back on to begin the growth of a thick ZnSe epilayer. All ZnSe growths were carried out at 200°C for 30 minutes under Se-rich condition [33, 34], except for one particular case, where the growth was carried out at 300°C. The resulting epilayers were nominally 230 nm thick. Light was sourced from a Xe lamp (Oriel 6255) operated at three different powers: 80 W, 115 W and 150 W. A 0.6 neutral density (ND) filter was placed between the lamp and the sample, and the beam was partially collimated to a spot size of approximately 25 cm$^2$ on the sample. The light flux was held constant throughout the duration of the growth.

*Characterization*

The layer thickness was confirmed by SEM image and ω-2θ HRXRD curve simulation. The estimated thicknesses of all ZnSe layers were approximately the same. AFM was used to measure the surface morphology of the ZnSe epilayers as a function of the Xe lamp power. Low temperature PL of the ZnSe epilayers was measured with a GaN diode laser (405 nm) operated at an excitation power ~1 mW and using a 435 nm long-pass filter to block the laser line. Photoluminescence of the underlying GaAs epilayers and ZnSe/GaAs interfaces were obtained with a Nd:YVO laser (532 nm) to avoid pumping the ZnSe epilayer. The excitation power was ~1.2 mW, and using a 570 nm long-pass filter. Both measurements were carried out with a 150 l/mm grating.

# References

1. Metze, G. M, Calawa, A. R. & Mavroides, J. G. An investigation of GaAs films grown by MBE at low substrate temperatures and growth rates. *J. Vac. Sci. Technol. B* **1**, 166-169 (1983).
2. Ahmed, S. *et al.* Use of nonstoichiometry to form GaAs tunnel junctions. *Appl. Phys. Lett.* **71**, 3667-3669 (1997).
3. Elman, B. *et al.* Low substrate temperature molecular beam epitaxial growth and the critical layer thickness of InGaAs Grown on GaAs. *J. Appl. Phys.* **70**, 2634-2640 (1991).
4. Oye, M. M. *et al.* Molecular-beam epitaxy growth of device-compatible GaAs on silicon substrates with thin (~80 nm) $Si_{1-x}Ge_x$ step-graded buffer layers for high-κ III-V metal-oxide-semiconductor field effect transistor applications. *J. Vac. Sci. Technol. B* **25**, 1098-1102 (2007).
5. Eyink, K. G. *et al.* A Comparison of the critical thickness for MBE grown LT-GaAs determined by in-situ ellipsometry and transmission electron microscopy. *J. Electron. Mater.* **26**, 391-396 (1997).
6. Park, K. *et al.* Unveiling interfaces between In-rich and Ga-rich GaInP vertical slabs of laterally composition modulated structures. *Appl. Phys. Express.* **10**, 025801 (2017) and references therein.
7. Cook, J. W. Jr., Eason, D. B., Vaudo, R. P. & Schetzina, J. F. Molecular-beam epitaxy of ZnS using an elemental S source. *J. Vac. Sci. Technol. B* **10**, 901-904 (1992).
8. Nakada, T. & Shirakata, S. Impacts of pulsed-laser assisted deposition on CIGS thin films and solar cells. *Sol. Energ. Mat. Sol.* **95**, 1463-1470 (2011).
9. Tu, C. W. *et al.* Laser-modified molecular beam epitaxial growth of (Al)GaAs on GaAs and $(Ca,Sr)F_2$/GaAs substrates. *Appl. Phys. Lett.* **52**, 966-968 (1988).
10. Donnelly, V. M. *et al.* Laser-assisted metalorganic molecular beam epitaxy of GaAs. *Appl. Phys. Lett.* **52**, 1065-1067 (1988).
11. Koestner, R. J., Liu, H. Y., Schaake, H. F. & Hanlon, T. R. Improved structural quality of molecular-beam epitaxy HgCdTe films. *J. Vac. Sci. Technol. A* **7**, 517-522 (1989).
12. Kitagawa, M. *et al.* Photo-assisted homoepitaxial growth of ZnS by molecular beam epitaxy. *J. Cryst. Growth* **101**, 52-55 (1990).

13. Giles, N. C. *et al.* Photoluminescence spectroscopy of CdTe grown by photoassisted MBE. *J. Cryst. Growth* **101**, 67-72 (1990).

14. Venkatasubramanian, R. *et al.* Incorporation processes in MBE growth of ZnSe. *J. Cryst. Growth* **95**, 533-537 (1989).

15. Bicknell, R. N., Giles, N. C., Schetzina, J. F. & Hitzman, C. Controlled substitutional doping of CdTe thin films grown by photoassisted molecular-beam epitaxy. *J. Vac. Sci. Technol. A* **5**, 3059-3063 (1987).

16. Walker, C. T. *et al.* Blue-green II-VI laser diodes. *Physica B* **185**, 27-35 (1993).

17. Fukada, T., Matsumura, N., Fukushima, Y. & Saraie, J. Low-temperature growth of ZnSe by photoassisted molecular beam epitaxy. *Jpn. J. Appl. Phys. Part 2* **29**, L1585-L1587 (1990).

18. Smathers, J. B., Kneedler, E., Bennett, B. R., & Jonker, B. T. Nanometer scale surface clustering on ZnSe epilayers. *Appl. Phys. Lett.* **72**, 1238-1240 (1998).

19. Marfaing, Y. Light-induced effects on the growth and doping of wide-bandgap II-VI compounds. *Semicond. Sci. Technol.* **6**, A60-A64 (1991).

20. Park, K., Beaton, D., Steirer, K. X. & Alberi, K. Effect of ZnSe/GaAs interface treatment in ZnSe quality control for optoelectronic device applications. *Appl. Surf. Sci.* **405**, 247-254 (2017).

21. Barabasi, A. L. & Stanley, H. E.; Fractal Concepts in Surface Growth (Cambridge University Press, 1995).

22. Beaton, D. A., Sanders, C. & Alberi, K. Effects of incident UV light on the surface morphology of MBE grown GaAs. *J. Cryst. Growth* **413**, 76-80 (2015).

23. Simpson, J. *et al.* Photoassisted molecular beam epitaxial growth of ZnSe under high UV irradiances. *Semicond. Sci. Technol.* **7**, 460-463 (1992).

24. Liu, Q. *et al.* The influence of structural defects in ZnSe/GaAs heterostructures on luminescence properties. *J. Phys. D: Appl. Phys.* **31**, 2421-2425 (1998) and reference therein.

25. Bourret, E. D., Zach, F. X., Yu, K. M. & Walker, J. M. Growth and characterization of ZnSe grown by organometallic vapor phase epitaxy using diisopropyl selenide and diethyl zinc. *J. Cryst. Growth* **147**, 47-54 (1995).

26. Matsumura, N. *et al.* Photo-assisted MBE growth of ZnSe on GaAs substrates. *J. Cryst. Growth* **111**, 787-791 (1991).

27. Thomas, A. E., Russell, G. J. & Woods, J. Self-activated emission in ZnS and ZnSe. *J. Phys. C: Solid Stat. Phys.* **17**, 6219-6228 (1984).

28. Sanders, C. E., Beaton, D. A., Reedy, R. C. & Alberi, K.; Fermi energy tuning with light to control doping profiles during epitaxy. *Appl. Phys. Lett.* **106**, 182105 (2015).

29. Seong, T. -Y., Kim, J. H., Chun Y. S. & Stringfellow G. B. Effects of V/III ratio on ordering and antiphase boundaries in GaInP layers. *Appl. Phys. Lett.* **70**, 3137-3139 (1997).

30. Dunstan, D. J., Nicholls, J. E., Cavenett, B. C. & Davis, J. J. Zinc vacancy-associated defects and donor-acceptor recombination in ZnSe. *J. Phys. C: Solid Stat. Phys.* **13**, 6409-6419 (1980).

31. Pashley, M. D. & Li, D. Control of the Fermi-level position on the GaAs(001) surface: Se Passivation. *J. Vac. Sci. Technol. A* **12**, 1848-1854 (1994).

32. Lu, F. *et al.* Interfacial properties of ZnSe/GaAs heterovalent interfaces. *J. Cryst. Growth* **184/185**, 183-187 (1998).

33. Zhang, Q., Shen, A., Kuskovsky, I. L. & Tamargo, M. C. Role of magnesium in band gap engineering of sub-monolayer type-II ZnTe quantum dots embedded in ZnSe. *J. Appl. Phys.* **110**, 034302 (2011).

34. Islam, S. K., Tamargo, M. C., Moug, R. & Lombardi, R. J. Surface-enhanced Raman scattering on a chemically etched ZnSe surface. *Phys. Chem. C* **117**, 23372-23373 (2013).

## Acknowledgement

We thank D. A. Beaton for his useful discussion about MBE growth. We acknowledge the financial support of the Department of Energy Office of Science, Basic Energy Sciences under contract DE-AC36-08GO28308. The U.S. Government retains and the publisher, by accepting the article for publication, acknowledges that the U.S. Government retains a nonexclusive, paid up, irrevocable, worldwide license to publish or reproduce the published form of this work, or allow others to do so, for U.S. Government purposes.

## Author contribution

K.P. conducted MBE growth and measurements. K.A. and K.P. designed experiments and analyzed measurement results. All authors reviewed the manuscript.

## Additional information

Competing financial interests: The authors declare no competing financial interests.

**Figure set**

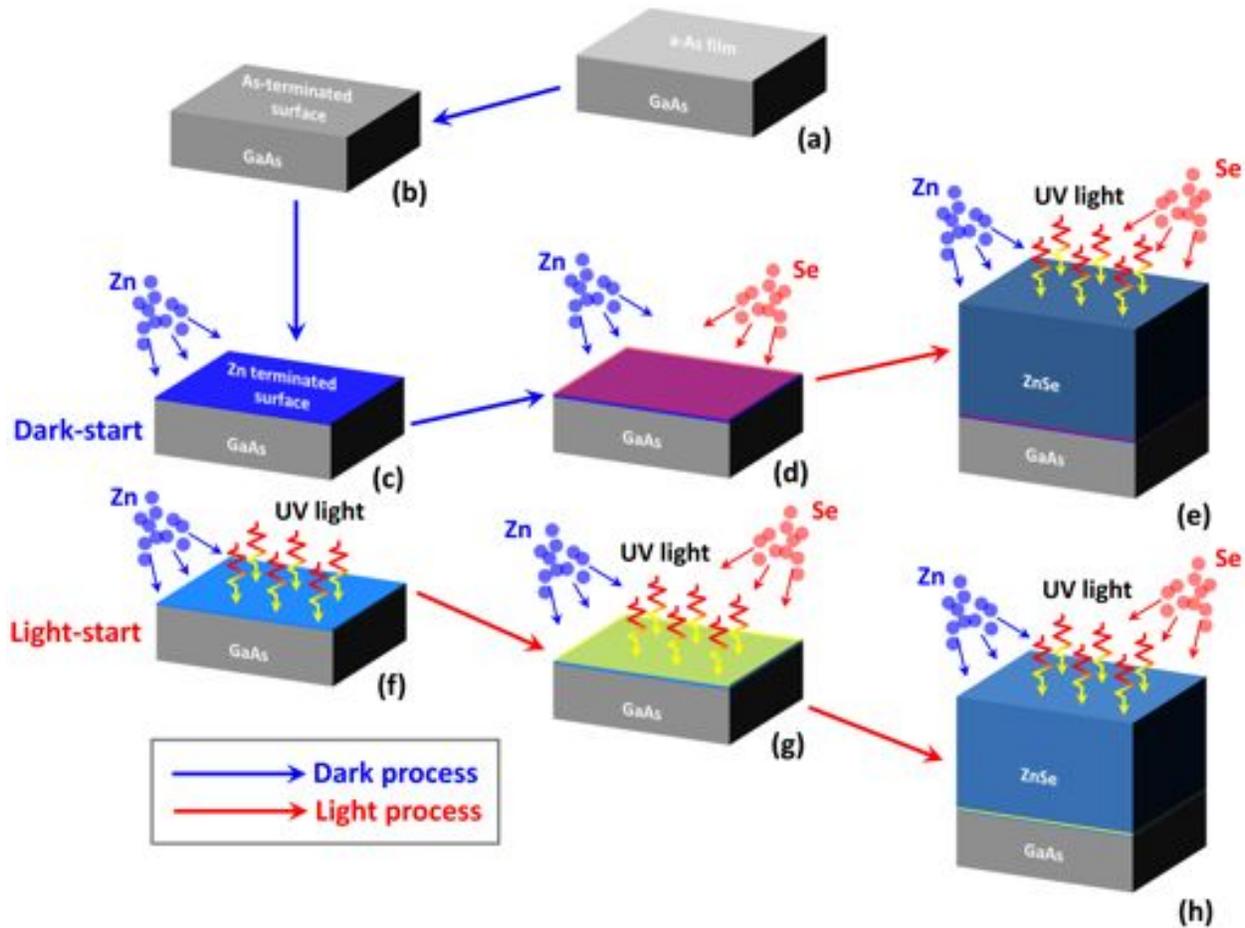

Figure 1. Schematic of the epitaxial growth procedure of ZnSe/GaAs heterostructure growth using either light-start or dark-start growth procedures. (a) Prior to ZnSe epilayer growth, the GaAs epilayer was grown in a dedicated III-V MBE chamber and was covered with an amorphous arsenic film prior to transferring it to a dedicated II-VI chamber. (b) The amorphous arsenic film was then thermally desorbed and the interface growth was initiated with a Zn pre-treatment under either a light-start or dark-start condition. During the light-start sample growth, UV light directed onto the growth surface from the beginning of Zn pre-treatment to the end of the ZnSe growth ((f), (g), (h)). During the dark-start sample growth, UV light was directed onto the growth surface only during ZnSe epilayer growth ((c), (d), (e)).

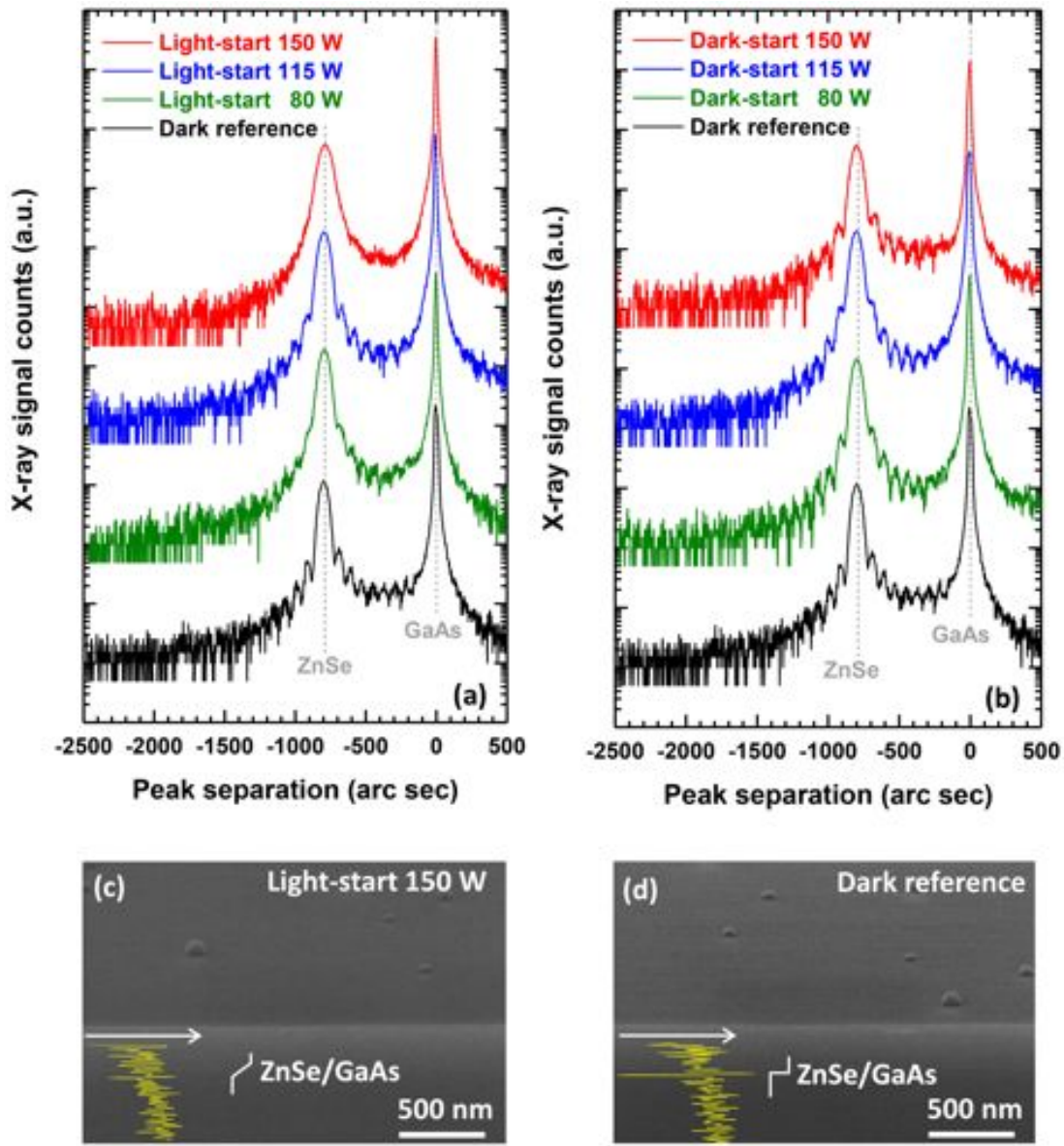

Figure 2. (a) and (b) present the HRXRD curves of samples grown under light-start and dark-start conditions, respectively. From bottom to top the Xe lamp power increased from 0 W to 150 W. Fringes between the GaAs and ZnSe peaks disappeared with increase of Xe lamp power in light-start samples, indicating diffusion at the interface. The fringes remained clear in all dark-start samples, indicating the presence of an abrupt interface. (c) and (d) present tilt-view SEM images of the light-start sample grown under a lamp power of 150 W as well as the dark reference sample. The yellow curves display a first derivative of image contrast profile at the ZnSe/GaAs interface. The contrast profile is sharp at the interface of the dark reference sample.

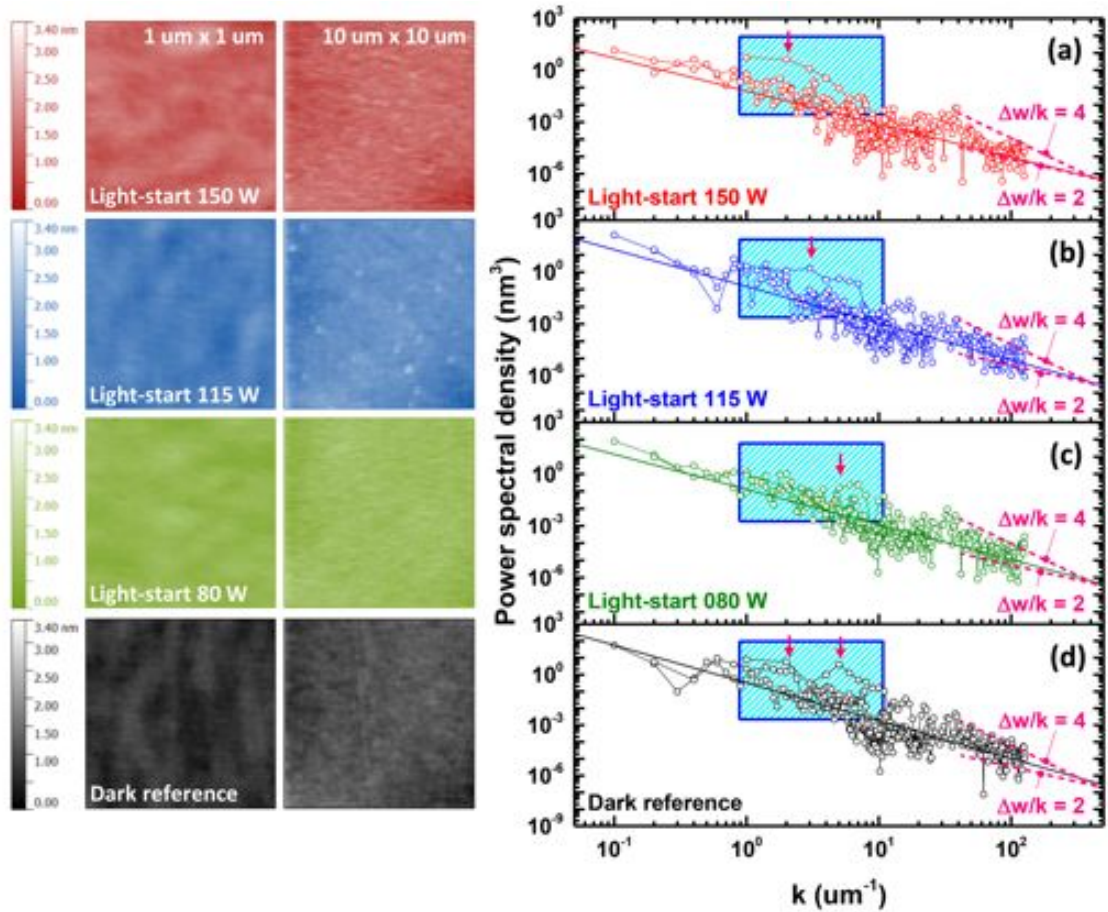

Figure 3. Left figure displays 1 × 1 μm$^2$ and 10 × 10 μm$^2$ surface AFM images of the light-samples and the dark reference sample. (a)-(d) displays their corresponding PSD curves. The curve slopes approached 2 with increasing Xe lamp power. Regions of elevated PDS shifted to lower spatial frequencies with higher lamp power, which indicates that the period of the hillocks on the sample surface became larger. The bright circular dots are in the AFM images are associated with SeO$_2$ islands.

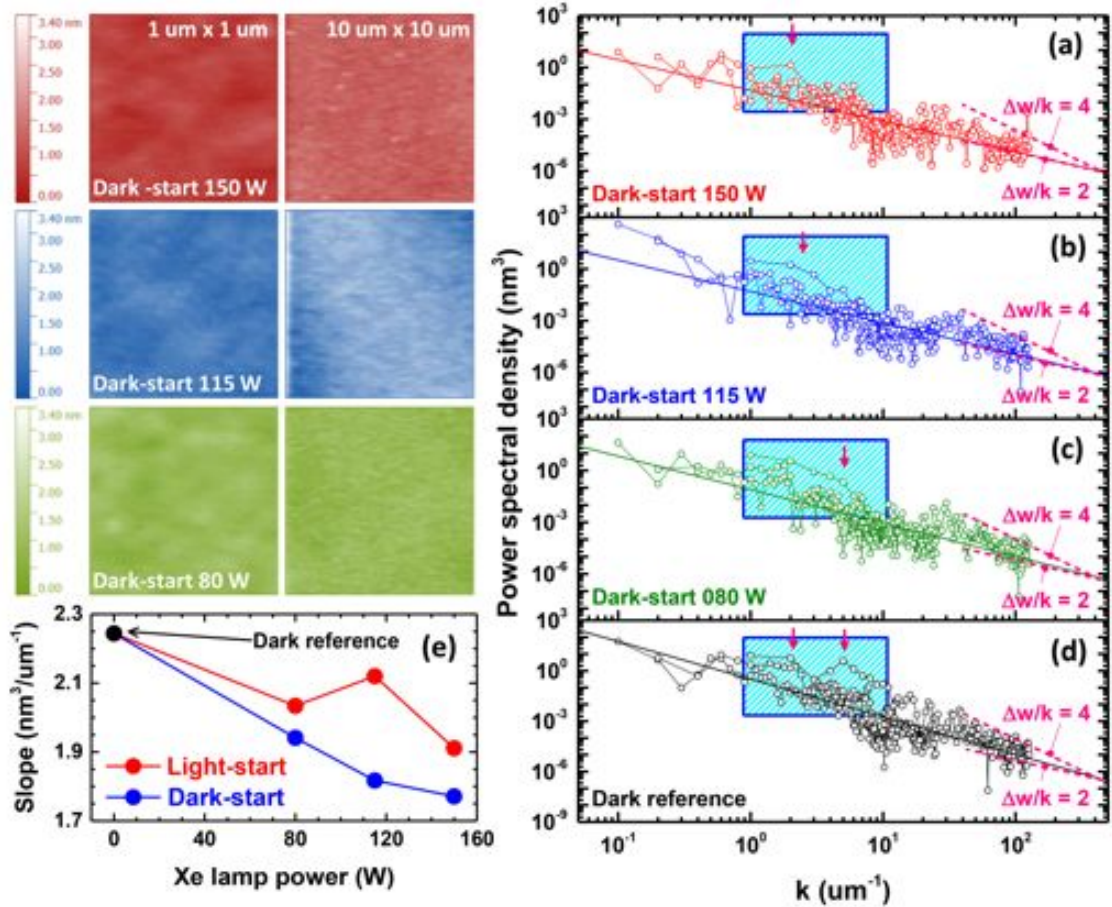

Figure 4. Left figure displays 1 × 1 μm$^2$ and 10 × 10 μm$^2$ surface AFM images of the dark-samples. (a)-(d) displays their corresponding PSD curves. The curve slopes approached 2 with increasing Xe lamp power. Regions of elevated PSD shifted to lower spatial frequency with increasing lamp power, which indicates that the period of the hillocks on the sample surface became larger. (e) presents the PSD slopes of the samples as a function of Xe lamp power. The PSD curve slopes of the light-start and dark-start samples decreased with increasing Xe lamp power.

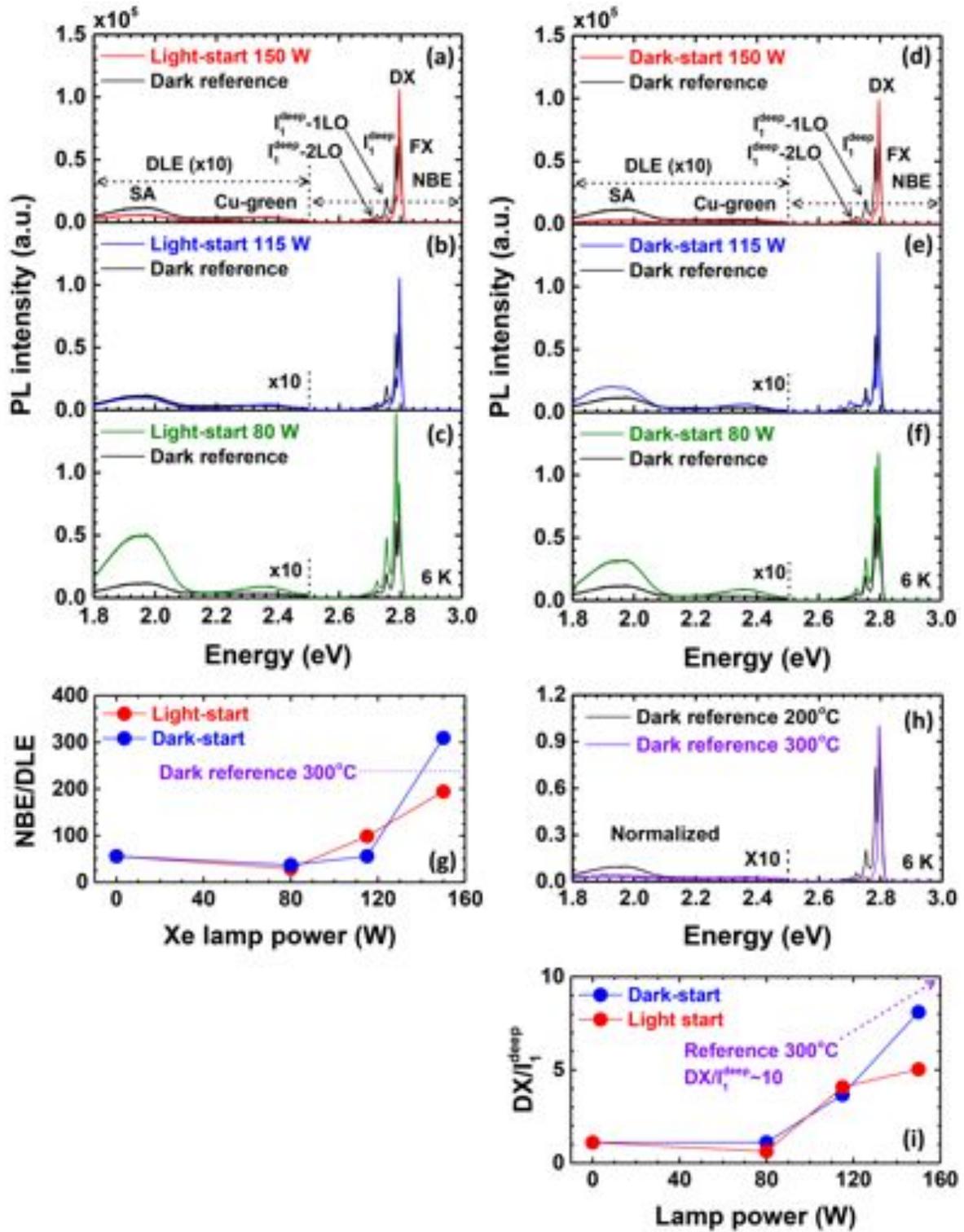

Figure 5. (a)-(c) and (d)-(f) displays PL emission from the ZnSe epilayers grown under light-start and dark-start conditions, respectively. The PL spectra consist emission from of NBE and DLE

regions. The DLE region is divided into SA and Cu-green emission. (g) presents the NBE/DLE peak intensity ratio of light-start samples as a function of Xe lamp power. (h) shows the normalized PL spectrum of dark reference samples grown at 200°C and 300°C. With increasing Xe lamp power, the PL spectrum of the 200°C-grown ZnSe epilayer approaches that of the 300°C grown ZnSe epilayer, with suppressed $I_1^{deep}$ and DLE emission. (i) presents $DX/I_1^{deep}$ of light-start and dark-start samples. $DX/I_1^{deep}$ increase with lamp power approaching the one of 300°C-grown ZnSe.

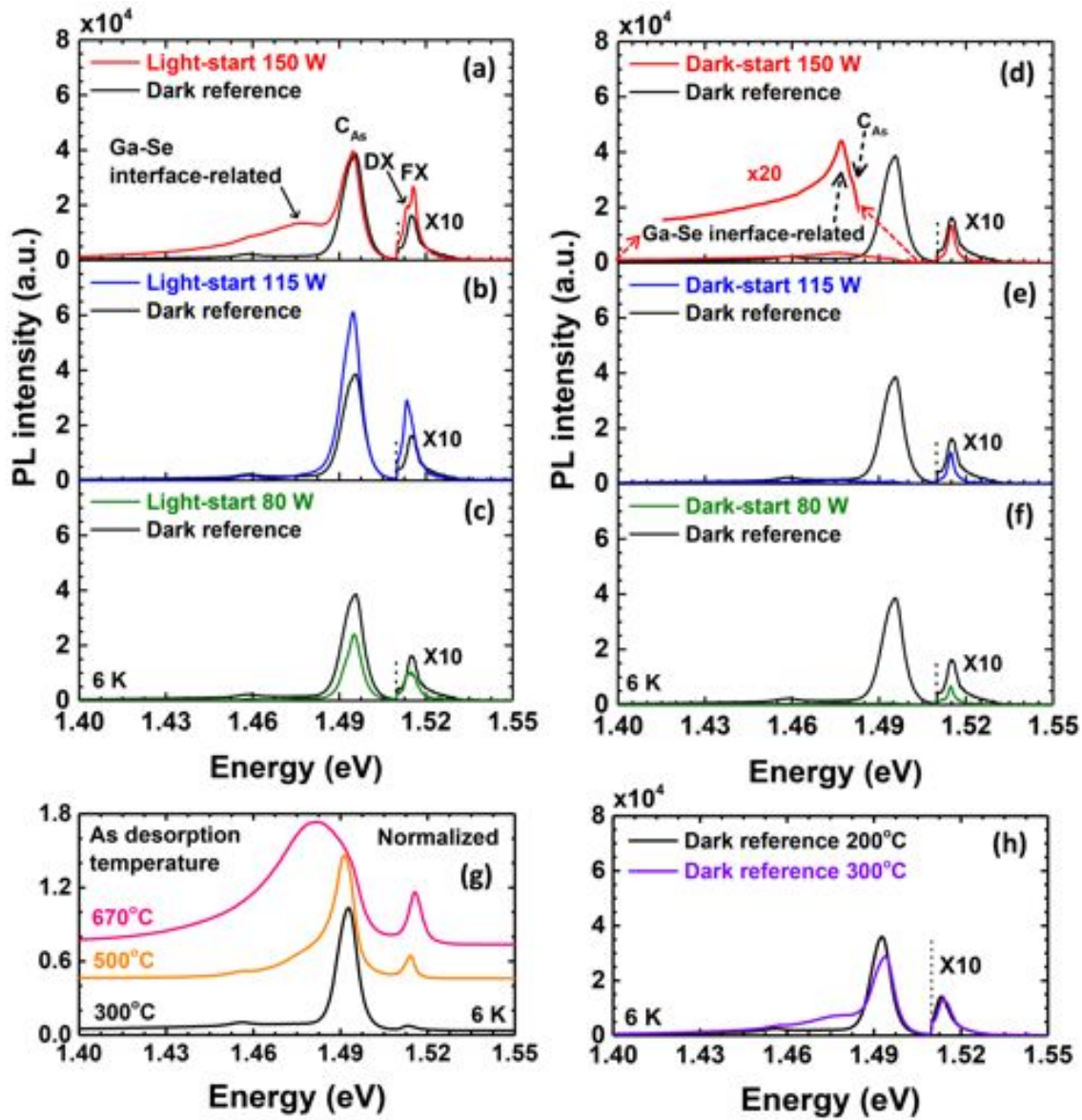

Figure 6. (a)-(c) and (d)-(f) displays PL emission from the ZnSe/GaAs interface and the GaAs of samples grown under light-start and dark-start conditions, respectively. From top to bottom, the Xe lamp power decreases from 150 W to 80 W. All spectra were compared with the dark reference sample. (g) shows the normalized PL spectra of the 200°C-grown ZnSe samples corresponding to As desorption temperatures increasing from 300°C to 670°C. The peak corresponding to Ga-Se bonding increases with increasing As desorption. (h) shows PL emissions of dark reference samples grown at 200°C and 300°C. The spectrum of the samples grown under the higher Xe lamp powers mimic that of the 300°C-grown sample.

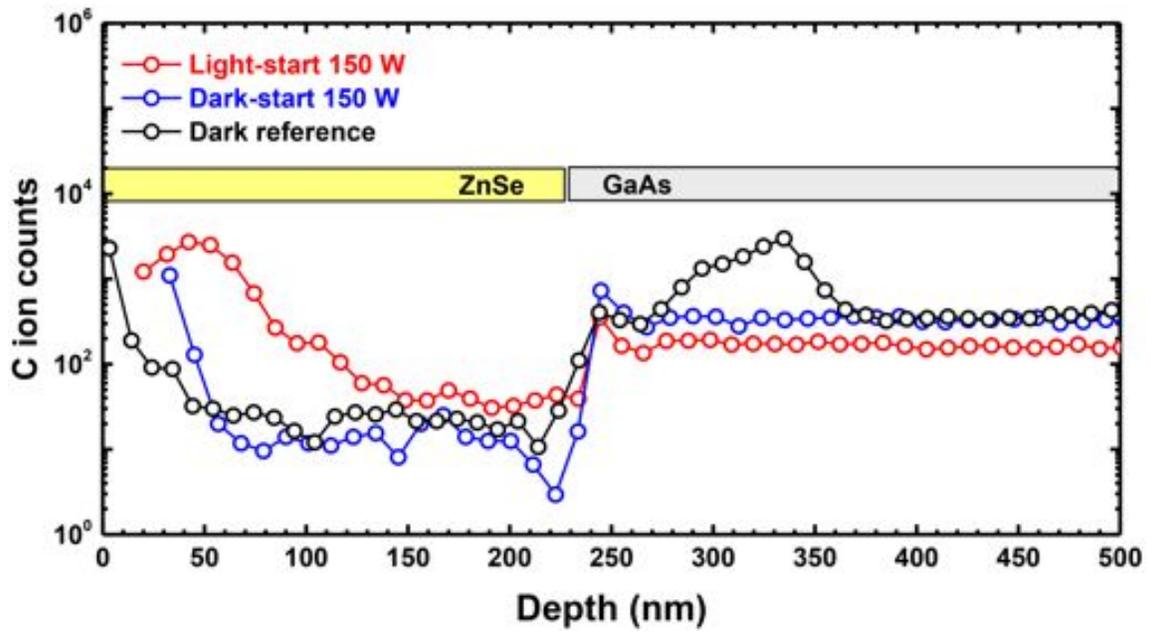

Figure 7. Cross-sectional carbon profile of light-start, dark-start and dark reference samples obtained by SIMS. For the light-stimulated samples, the Xe lamp power was 150 W. There is no significant difference among the samples in cross-sectional carbon ion counts.